\documentclass[reprint,superscriptaddress,amsmath,amssymb,aps,pra]{revtex4-1}
\usepackage[latin9]{inputenc}
\usepackage{graphicx}
\usepackage{xcolor}
\usepackage{amsmath}
\usepackage{amssymb}
\usepackage{esint}

\makeatletter

\usepackage{dcolumn}% Align table columns on decimal point
\usepackage{bm}% bold math
\makeatother

\begin{document}

\title{Cavity magnomechanics}

\date{\today}

\author{Xufeng Zhang}

\address{Department of Electrical Engineering, Yale University, New Haven,
Connecticut 06511, USA}

\author{Chang-Ling Zou}

\address{Department of Electrical Engineering, Yale University, New Haven,
Connecticut 06511, USA}

\address{Department of Applied Physics, Yale University, New Haven, Connecticut
06511, USA}

\author{Liang Jiang}

\address{Department of Applied Physics, Yale University, New Haven, Connecticut
06511, USA}

\author{Hong X. Tang}\thanks{To whom correspondence should be addressed. E-mail: hong.tang@yale.edu}

\address{Department of Electrical Engineering, Yale University, New Haven,
Connecticut 06511, USA}

\address{Department of Applied Physics, Yale University, New Haven, Connecticut
06511, USA}

\maketitle \textbf{A dielectric body couples with
electromagnetic fields through radiation pressure
and electrostrictive forces, which mediate
phonon-photon coupling in cavity optomechanics
\cite{Aspelmeyer2014}. In a magnetic medium,
according to Korteweg-Helmholtz formula
\cite{Zahn2006}, magnetostrictive forces should
arise and lead to phonon-magnon interaction. Here
we report such a coupled phonon-magnon system
based on ferrimagnetic spheres, which we term as
cavity magnomechanics, by analogy to cavity
optomechanics. Coherent phonon-magnon
interactions, including electromagnetically
induced transparency and absorption, are
demonstrated. Excitingly, due to strong
hybridization of magnon and microwave photon
modes and their high tunability, our platform
exhibits new features including parametric
amplification of magnons and phonons, triply
resonant photon-magnon-phonon coupling and phonon
lasing. Our work demonstrates the fundamental
principle of cavity magnomechanics and its
application as a new information transduction
platform based on coherent coupling between
photons, phonons and magnons. }

Mechanical oscillators have been recently widely
studied as a transducer mediating the coherent
signal conversion between different systems
\cite{Aspelmeyer2014}. Particularly, radiation
force \cite{Li2008, Park2009, Weis2010,
Safavi-Naeini2011, Hill2012}, electrostatic force
\cite{Teufel2011, Andrews2014_NPhys, Bagci2014}
and piezoelectric force \cite{Bochmann2013,
Fan2015} have been utilized for coupling phonon
with optical or microwave photons. Such
interaction mechanisms lead to the fast
development of a variety of cavity electro- and
opto-mechanical systems \cite{Aspelmeyer2014},
but they all intrinsically lack good tunability.
The magnetostrictive force \cite{Zahn2006}
provides an alternative mechanism to allow a
different information carrier -- magnon -- to
couple with phonon. Magnon is a collective
excitation of magnetization, whose frequency can
be tuned at will by adjusting bias magnetic field
\cite{Serga2010, Lenk2011, Chumak2015}. The
magnetostrictive interaction has been long
overlooked for information processing as it is
negligibly weak in commonly used dielectric or
metallic materials. However, in magnetic
materials such as yttrium iron garnet (YIG,
Y$_3$Fe$_5$O$_{12}$), the magnetostrictive force
becomes dominant, which provides a great
opportunity to establish an highly tunable hybrid
system for coherent information processing.
Thanks to the excellent material property of YIG,
the magnomechanical system can be further
integrated with opto- or electro-mechanical
elements, providing an excellent platform for
quantum state transfer among different physical
systems.

Here, we demonstrate an intriguing cavity
magnomechanical system in which magnon couples
with phonon through magnetostrictive interaction,
resulting in hallmark coherent phenomena such as
magnomechanically induced transparency/absorption
(MMIT/MMIA) and magnomechanical parametric
amplification (MMPA). During such processes,
magnons are in the hybridized state with cavity
microwave photons as they are strongly coupled to
each other \cite{Tabuchi2014_PRL, Zhang2014_PRL,
Goryachev2014, Bai2015, Kurizki2015}. Therefore
coherent signal conversions among these three
different information carriers are realized in a
single device. The magnetic field dependence of
magnon provides our system with unprecedented
tunability compared with opto- or
electro-mechanical systems. Moreover, the great
flexibility of this system allows us to achieve
triple-resonance among magnon, phonon and photon,
which drastically enhances the magnomechanical
interaction. The principles demonstrated in our
room temperature experiments can be readily
applied to the quantum regime at millikelvin
temperature, opening up great opportunities in
various applications, such as tunable microwave
filter and amplifier \cite{Bergeal2010},
long-lifetime quantum memories
\cite{Fiore2011_PRL}, microwave-to-optics
conversion \cite{Andrews2014_NPhys}.

\begin{figure*}
\centering
\includegraphics[width=183mm]{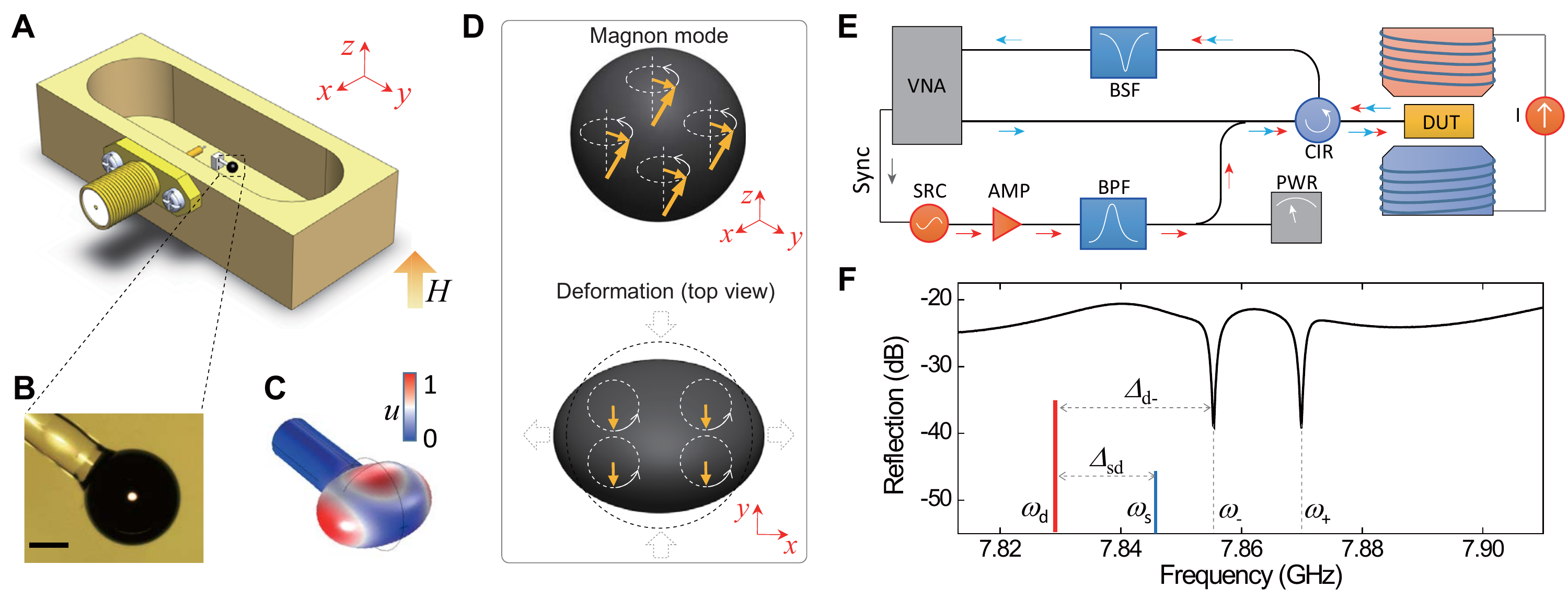}
\protect\protect\protect\protect\protect\caption{\textbf{Device
schematic and measurement setup.} \textbf{(A)}
Schematic of the device that consists of a
three-dimensional copper cavity (only bottom half
is shown) and a YIG sphere. The YIG sphere is
placed near the maximum microwave magnetic field
(along $y$ direction) of the cavity TE$_{011}$
mode. A uniform external magnetic field ($H$) is
applied along $z$ direction to bias the YIG
sphere for magnon-photon coupling. \textbf{(B)}
Optical image of the highly polished
250-$\mu$m-diameter YIG sphere that is glued to a
125-$\mu$m-diameter supporting silica fiber. The
gluing area is minimized to reduce the contact
damping to the phonon mode. Scale bar: 100
$\mu$m. \textbf{(C)} Simulated mechanical
displacement ($u$) of the S$_{1,2,2}$ phonon mode
in the YIG sphere which has the strongest
magnomechanical interaction with the uniform
magnon mode. \textbf{(D)} An intuitive
illustration of the magnomechanical coupling. Top
panel shows the uniform magnon excitation in the
YIG sphere. Bottom panel shows that the
magnon-induced magnetization (vertical yellow
arrows) causes the deformation (compression along
$y$ direction) of the YIG sphere (and vise
versa). \textbf{(E)} Schematic illustration of
the measurement setup. VNA: vector network
analyzer; SRC: microwave source for driving; AMP:
microwave amplifier; BPF: bandpass filter; PWR:
microwave power meter; CIR: circulator; BSF:
bandstop filter; DUT: device-under-test.
\textbf{(F)} Black curve: cavity reflection
spectrum when magnon is on-resonance with the
cavity photon mode. The two dips represent the
maximum hybridized modes
$\hat{A}_{\pm}=\frac{1}{\sqrt{2}}(\hat{a}\pm\hat{m})$.
Red and blue vertical lines indicate the applied
drive and probe, respectively. The probe is swept
across the hybrid mode resonance.
$\Delta_\mathrm{sd}$: two photon (probe-drive)
detuning; $\Delta_{\mathrm{d}-}$: drive-resonance
detuning.}

\label{fig:fig1}
\end{figure*}

\vbox{}
\noindent\textbf{Magnetostrictive interaction}\\
The device used in our experiments is
schematically shown in Fig.\,\ref{fig:fig1}A. The
key component is a highly polished single crystal
YIG sphere glued to the end of a silica fiber for
supporting purpose (Fig.\,\ref{fig:fig1}B). With
an external magnetic field $H$ biased along $z$
direction, a uniform magnon mode resonates in the
YIG sphere at frequency
$\omega_{\mathrm{m}}=\gamma H$, where $\gamma$ is
the gyromagnetic ratio. The YIG sphere is also an
excellent mechanical resonator
(Fig.\,\ref{fig:fig1}C) thanks to its superior
material and geometrical properties. The varying
magnetization induced by the magnon excitation
inside the YIG sphere causes deformation of its
spherical geometry (and vise versa), introducing
the coupling between magnon and phonon modes
(Fig.\,\ref{fig:fig1}D). Considering the large
frequency mismatch between the magnon and the
phonon modes (gigahertz v.s. megahertz) with our
experiment parameters, a strong parametric drive
is used to compensate their frequency difference.
In this case, the system is described by an
radiation pressure-like, dispersive interaction
Hamiltonian $\mathcal{H}=\hbar
g_{\mathrm{mb}}\hat{m}^{\dagger}\hat{m}(\hat{b}+\hat{b}^{\dagger})$,
where $\hbar$ is the reduced Planck's constant,
$\hat{b}$ ($\hat{m}$) is the boson operator of
the phonon (magnon) mode, and $g_{\mathrm{mb}}$
is the single magnon-phonon coupling strength.

\begin{figure*}
\centering
\includegraphics[width=183mm]{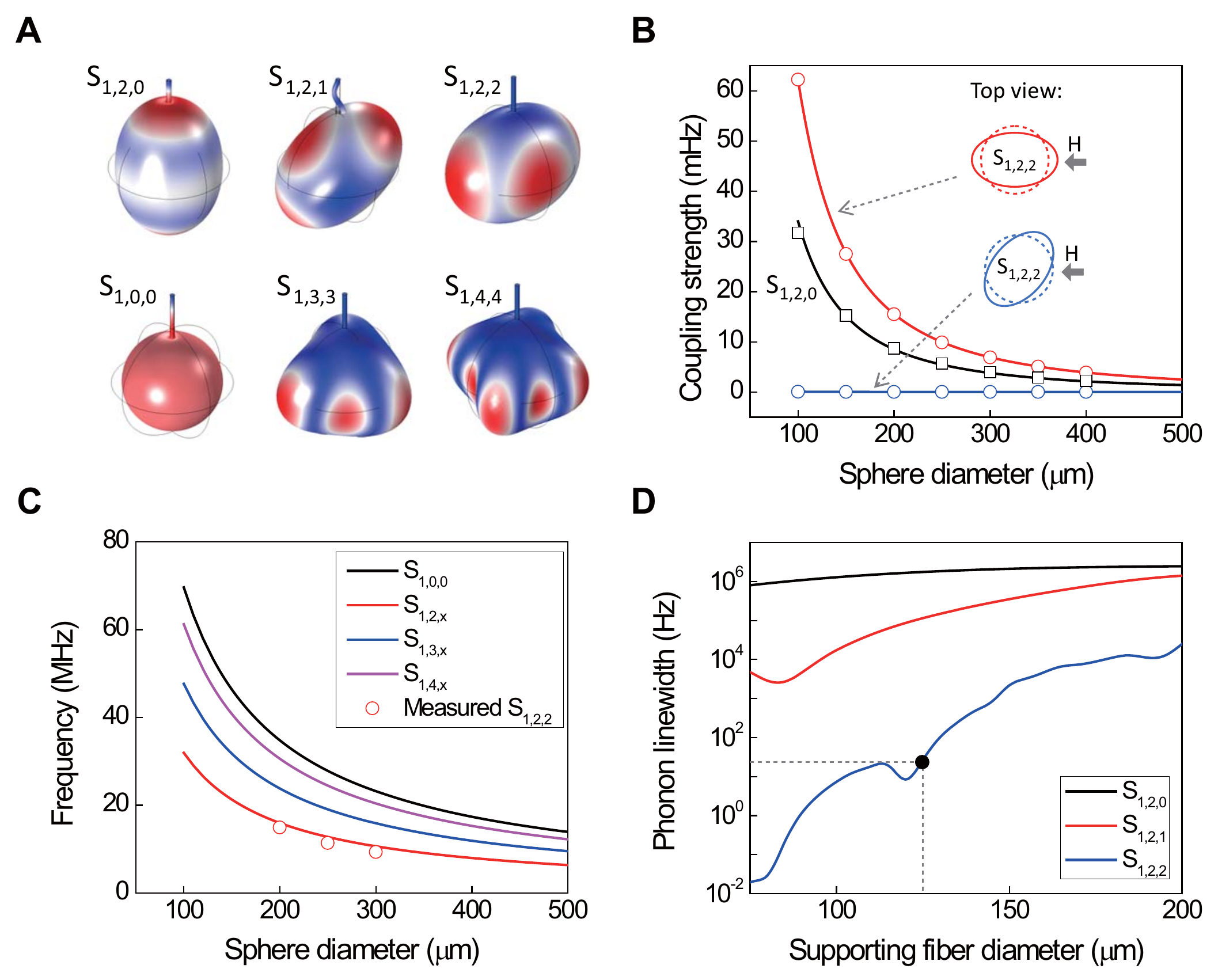}
\protect\protect\protect
\protect\protect\caption{\textbf{Analysis of the
phonon modes and magnetostrictive coupling
strengths.} \textbf{(A)} Simulated displacement
profiles of the low order phonon modes in the YIG
sphere (with a small supporting fiber).
S$_{1,l,m_a}$ represents the spheroidal mode with
a radial mode number of 1, an angular mode number
of $l$, and an azimuthal mode number of $m_a$.
Only one of the $2l+1$ degenerate modes is
plotted for each $l$. \textbf{(B)} Theoretical
prediction of the magnomechanical coupling
strength as a function of YIG sphere diameter for
the S$_{1,2,0}$ (black) and S$_{1,2,2}$ modes
(red and blue, corresponding to different bias
field directions). Solid lines are numerical
calculations while symbols are analytical
fittings. \textbf{(C)} Phonon mode frequency as a
function of the YIG sphere diameter. Solid lines
are the theoretical calculations, showing an
inverse proportional dependence, while red
circles are the measurement results. \textbf{(D)}
Simulated phonon linewidth due to clamping loss
as a function of the supporting fiber diameter
for the S$_{1,2,m_a}$ modes. Black dot indicates
the experiment parameter, showing an
anchor-loss-limited linewidth of 20 Hz.}

\label{fig:fig2}
\end{figure*}

\vbox{}
\noindent\textbf{Spheroidal phonon modes}\\
The magnetostrictive coupling strength is
determined by the mode overlap between the
uniform magnon mode and the phonon modes. In a
YIG sphere, there exist various phonon modes,
each with a different displacement profile and
consequently a different coupling strength with
the magnon mode. Figure\,\ref{fig:fig2}A plots
the typical profiles of the lowest order
spheroidal phonon modes S$_{1,l,m_{a}}$ ($l$ and
$m_{a}$ are the angular and azimuthal mode
numbers, respectively), among which the
S$_{1,2,2}$ mode shows the highest coupling
strength when the bias field is along the
direction of maximum displacement
(Fig.\,\ref{fig:fig2}B). Therefore in our
experiments we focus only on the S$_{1,2,2}$
mode. Although a YIG sphere with a smaller
diameter is favorable for achieving larger
coupling strengths (Fig.\,\ref{fig:fig2}B), it
also results in a higher frequency for the phonon
mode (Fig.\,\ref{fig:fig2}C), which in turn leads
to lower responsivity to the parametric drive, so
a trade-off has to be made when choosing the
sphere size. In our experiments, a
250-$\mu$m-diameter YIG sphere is used,
corresponding to a phonon frequency
$\omega_{\mathrm{b}}/2\pi=11.42$ MHz and a
coupling strength $g_{\mathrm{mb}}/2\pi\leq9.9$
mHz. With an external drive of 0 dBm, the linear
magnon-phonon coupling can be enhanced to around
$30$ kHz, which is two orders of magnitude larger
than the phonon dissipation rate
$\kappa_\mathrm{b}$.

Magnetostriction mediates the coupling between
magnons and photons. However, in order to achieve
coherent magnon-phonon coupling, it is further
required that phonon mode should have relatively
long lifetime. Single crystal YIG has a garnet
structure that is known to exhibit very low
mechanical damping and therefore supports a
material-limited phonon lifetime over a
millisecond \cite{LeCraw1961}. The supporting
fiber that is glued to the YIG sphere reduces the
phonon lifetime (Fig.\,\ref{fig:fig2}D). In our
experiments, the measured linewidth of
S$_{1,2,2}$ phonon mode with a
125-$\mu$m-diameter supporting fiber is
$2\kappa_{\mathrm{b}}/2\pi=300$ Hz, which is
sufficiently small for observing coherent
magnon-phonon coupling phenomena.

\vbox{}
\noindent\textbf{Coherent magnomechanical interaction}\\
Figure\,\ref{fig:fig1}E plots the schematics of
our measurement setup at room temperature ambient
condition. The YIG sphere is placed inside a
three-dimensional microwave cavity
(Fig.\,\ref{fig:fig1}A). A weak probe signal is
sent into the cavity through a coaxial probe, and
by sweeping its frequency $\omega_{\mathrm{s}}$
and measuring the reflection, we can infer the
interaction among photon, magnon and phonon
inside the cavity. The YIG sphere is positioned
at the maximum microwave magnetic field of the
cavity TE$_{011}$ mode, which resonates at
$\omega_{\mathrm{a}}/2\pi=7.86$ GHz. By
controlling the bias magnetic field, we tune the
magnon close to resonance with the cavity photon
mode. This leads to the hybridization between
magnon and photon \cite{Zhang2014_PRL,
Tabuchi2014_PRL, Goryachev2014, Bai2015}, which
shows up in the reflection spectrum as a pair of
split normal modes (Fig.\,\ref{fig:fig1}F).
Because each of the two hybrid modes contains
magnon components, it coherently couples with the
phonon modes when the cavity is parametrically
driven by a strong microwave signal at
$\omega_{\mathrm{d}}$.

We first study the coherent magnomechanical
coupling for each individual hybrid mode by
applying an off-resonance microwave drive. In
this case, the cavity magnomechanical system can
be described by
\begin{eqnarray}
\mathcal{H}_{\mathrm{mb}} & = & \hbar g_{\mathrm{mb}}(\hat{b}+\hat{b}^{\dagger})(\cos^{2}\theta\hat{A}_{+}^{\dagger}\hat{A}_{+}+\sin^{2}\theta\hat{A}_{-}^{\dagger}\hat{A}_{-}),
\end{eqnarray}
where the two hybrid modes interact with the
phonon mode independently. Here,
$\hat{A}_{+}=\cos\theta\hat{a}+\sin\theta\hat{m}$
and
$\hat{A}_{-}=-\sin\theta\hat{a}+\cos\theta\hat{m}$
are quantized boson operators for hybridized
excitations constituted by magnon and microwave
photon ($\hat{a}$), with
$\theta=\frac{1}{2}\arctan\frac{2g_{\mathrm{ma}}}{\Delta_{\mathrm{ma}}}$
varies with photon-magnon coupling strength
$g_\mathrm{ma}$ and photon-magnon detuning
$\Delta_\mathrm{ma}=\omega_\mathrm{m}-\omega_\mathrm{a}$.
In our system, both the magnon and the cavity
photon modes have a relatively narrow linewidth
($2\kappa_{\mathrm{m}}/2\pi=1.12$ MHz and
$2\kappa_{\mathrm{a}}/2\pi=3.35$ MHz). As a
result, the hybrid mode linewidth is well below
the phonon frequency, leading our system deep
inside the resolved sideband regime, by analogy
with optomechanical systems
\cite{Aspelmeyer2014}. In this case, the
nonlinear interaction can be converted either
into the linear beam splitter model
$\hbar(G_{\mathrm{\pm}}\hat{A}_{\pm}^{\dagger}\hat{b}+G_{\mathrm{\pm}}^{*}\hat{A}_{\pm}\hat{b}^{\dagger})$
or the parametric oscillator model
$\hbar(G_{\mathrm{\pm}}\hat{A}_{\pm}^{\dagger}\hat{b}^{\dagger}+G_{\mathrm{\pm}}^{*}\hat{A}_{\pm}\hat{b})$
with the presence of an external drive, where
$G_{\mathrm{\pm}}=A_{\pm,\mathrm{ss}}g_{\mathrm{mb}}(1\mp
\mathrm{cos}2\theta)/2$ is the enhanced coupling
strength. Here, $A_{\pm,\mathrm{ss}}$ is the
steady state amplitude of the hybrid mode,
corresponding to the effective pumping of the
microwave drive on magnon due to the
magnon-photon hybridization.

\begin{figure*}
\centering
\includegraphics[width=183mm]{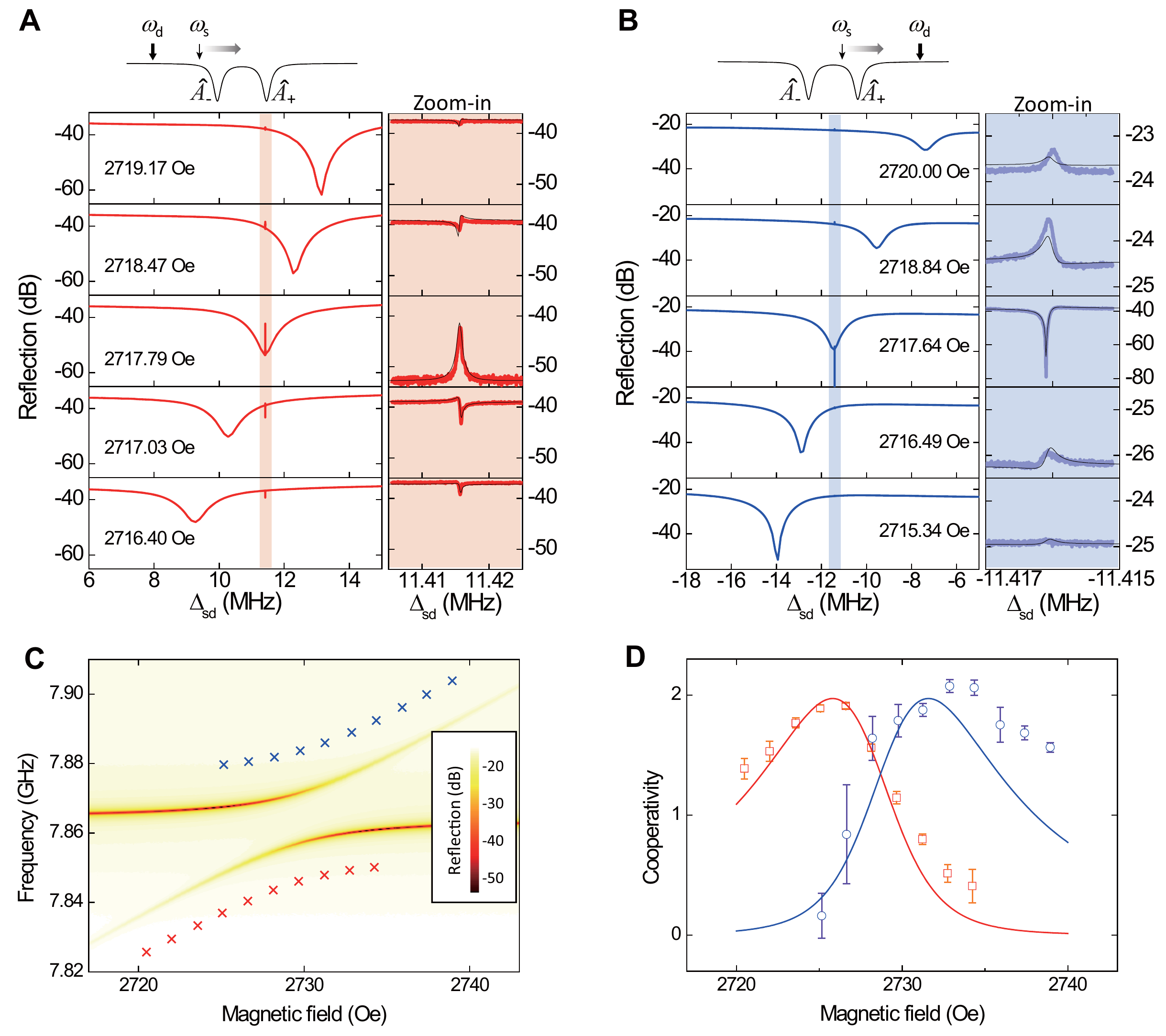}
\protect\protect\protect\protect\protect\caption{
\textbf{Tunable magnomechanically induced
transparency/absorption}. \textbf{(A)} Measured
reflection spectra near the lower hybrid mode
$\hat{A}_-$ as a function of the two-photon
detuning $\Delta_{\mathrm{sd}}$ for a series of
different bias magnetic fields. The broad dip
corresponds to the lower hybrid mode resonance,
whose lineshape changes with bias magnetic field
because of the change in the ratio between magnon
and photon components. A strong (26 dBm)
microwave drive is red-detuned with a fixed
driving frequency $\omega_{\mathrm{d}}$. A
Fano-like narrow resonance line shows up inside
the hybrid mode, which turns into a Lorentzian
transparency peak when
$\Delta_{\mathrm{d-}}=-\omega_{\mathrm{b}}$.
Zoom-in shows detailed spectra of the
magnomechanically induced resonances (shaded area
in (A)). \textbf{(B)} Measured reflection spectra
near the upper hybrid mode $\hat{A}_+$ with a
blue-detuned strong drive (22 dBm) for various
bias magnetic fields. When
$\Delta_{\mathrm{d+}}=\omega_{\mathrm{b}}$, the
magnomechanically induced narrow resonance shows
up as a Lorentzian absorption dip. Zoom-in shows
detailed spectra of the shaded area in (B).
\textbf{(C)} Reflection spectra of the hybrid
magnon-photon modes at various bias magnetic
fields. The crosses indicate the drive frequency
and bias magnetic field used for each data point
in (D). \textbf{(D)} The magnomechanical
cooperativity as a function of bias magnetic
field. For each measurement, the microwave drive
is detuned from the hybrid mode by
$\Delta_{\mathrm{d\pm}}=\pm\omega_{\mathrm{b}}$,
while the probe is swept across the hybrid mode
resonance. Red squares (blue circles) are for the
red (blue) detuning situation. Solid lines in (D)
and in the zoom-in of (A) and (B) are theoretical
calculations using only a single fitting
parameter $g_{\mathrm{mb}}$.}

\label{fig:fig3}
\end{figure*}

Figures\,\ref{fig:fig3}A and B plot the measured
reflection spectra for a series of bias magnetic
fields with a microwave drive at a fixed
frequency $\omega_{\mathrm{d}}$. To avoid the
influence of the other hybrid mode, the driving
signal is red (blue) detuned for the lower
(upper) hybrid mode, as illustrated by the top
insets. For the red-detuned drive, the power is
held constant at 26 dBm. In the spectra, the
broad Lorentzian-shaped resonance dip corresponds
to the hybrid mode $\hat{A}_{-}$, while the very
sharp modification of the spectra at the
two-photon detuning
$\Delta_{\mathrm{sd}}=\omega_{\mathrm{b}}$ is
evidence of coherent magnomechanical interaction.
The zoomed-in spectra in Fig.\,\ref{fig:fig3}A
show that these phonon-induced resonances have a
Fano-like shape that varies with bias magnetic
field. When the drive-resonance detuning
$\Delta_{\mathrm{d-}}=\omega_{\mathrm{d}}-\omega_{-}=-\omega_{\mathrm{b}}$,
the Fano-like resonance changes into a symmetric
Lorentzian-shaped transparency peak (MMIT). In
contrast, the Fano-like resonances in the spectra
for the blue-detuned drive (with a constant power
of 22 dBm) show an opposite symmetry
(Fig.\,\ref{fig:fig3}B). When the drive is blue
detuned to
$\Delta_{\mathrm{d+}}=\omega_{\mathrm{d}}-\omega_{+}=\omega_{\mathrm{b}}$,
such a resonance becomes a Lorentzian-shaped
absorption dip (MMIA).

One distinct advantage of magnon is that its
frequency is determined by the external bias
magnetic field and therefore can be conveniently
tuned in a broad range. By varying
$\Delta_{\mathrm{ma}}$, the percentage of magnon
component in the hybrid mode changes. Therefore,
the hybrid mode experiences different effective
dissipation rate, external coupling rate, as well
as effective coupling strength with the phonon
mode. As a result, the coherent magnomechanical
interaction is magnetically controllable, which
can be quantified by the dependence of the
cooperativity
$C=G_{\pm}^{2}/\kappa_{\pm}\kappa_{\mathrm{b}}$
on the bias magnetic field. The measured $C$-$H$
relation is plotted in Fig.\,\ref{fig:fig3}D. For
each measurement under a specific bias condition,
the drive frequency is detuned from the hybrid
mode by
$\Delta_{\mathrm{d}\pm}=\pm\omega_{\mathrm{b}}$,
as indicated by the crosses in
Fig.\,\ref{fig:fig3}C, while the driving power is
fixed constant at $30\,\mathrm{dBm}$. We can see
there exists an optimal condition for a maximum
$C$, as a the result of the competition between
the magnon and photon components in the hybrid
mode: more magnon component yields stronger
magnetostrictive coupling, while more photon
component leads to a higher driving efficiency.
From these measurement results we can extract the
magnon-phonon coupling strength
$g_{\mathrm{mb}}/2\pi=4.1$ mHz, in accordance
with our theoretical prediction
(Fig.\,\ref{fig:fig2}B).

\begin{figure*}
\centering
\includegraphics[width=183mm]{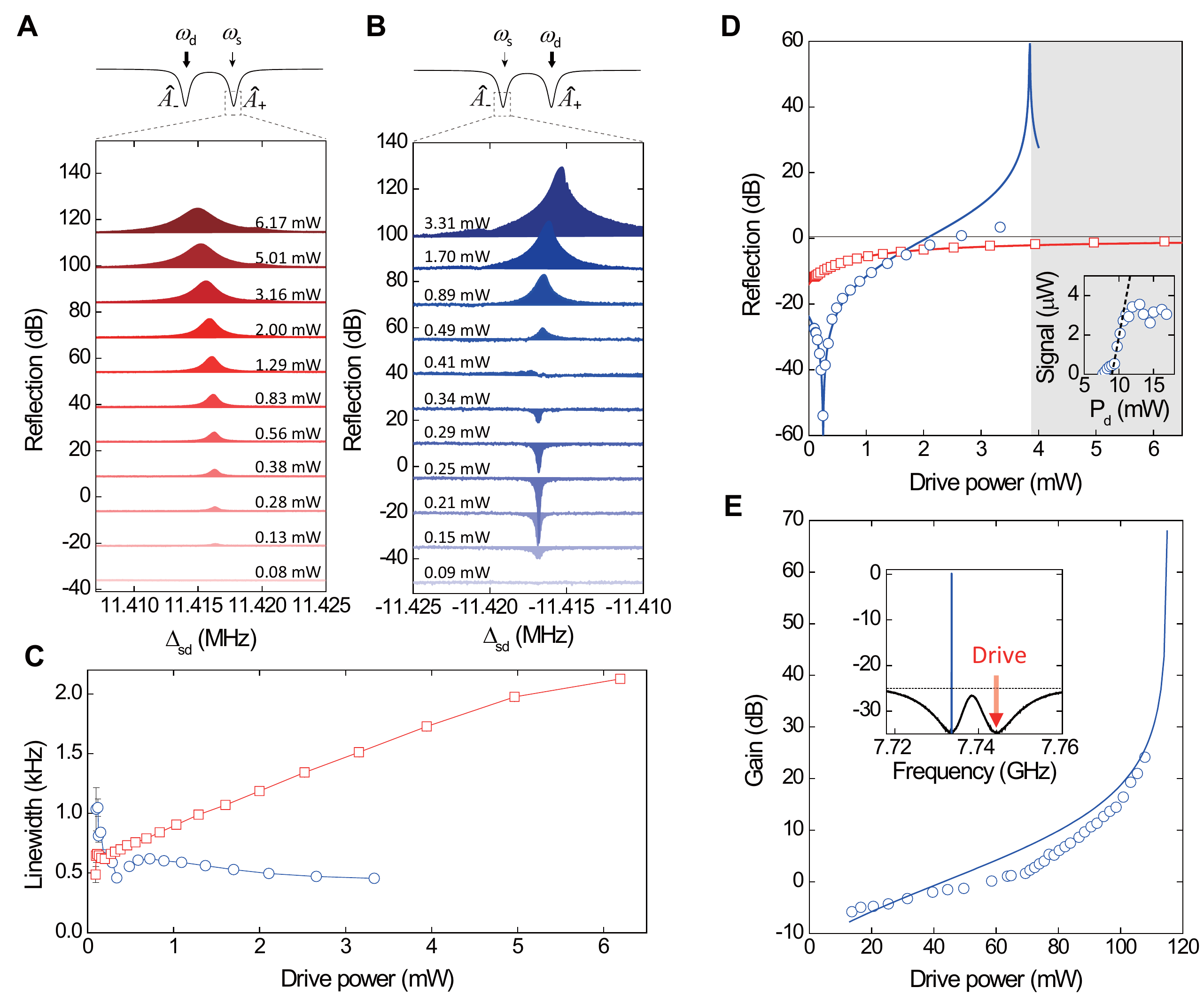}
\protect\protect\protect\caption{\textbf{Enhanced
magnomechanical coupling in the triply resonant
system.} \textbf{(A)} Magnomechanically induced
transparency (MMIT) signal for a red-detuned
drive at various driving powers. \textbf{(B)}
Magnomechanically induced absorption (MMIA) and
magnomechanical parametric gain (MMPA) signal for
a blue-detuned drive at various driving powers.
\textbf{(C)} The linewidth of the
magnomechanically induced resonance as a function
of the drive power. \textbf{(D)}
Magnomechanical-interaction-modified on-resonance
reflectivity of the hybrid mode as a function of
the drive power. Shaded area indicates the
instable regime. Inset: Measured power of the
Stokes sideband of the driving signal. The
threshold behavior indicates the onset of phonon
lasing. \textbf{(E)} Magnomechanical parametric
gain as a function of the drive power in an
over-coupled hybrid system. Inset: Measured
reflection spectrum that shows a 25-dB gain. In
the main panels of (C)--(E), blue circles (red
squares) are for the blue (red) detuning, and
solid lines are theoretical calculations.}

\label{fig:fig4}
\end{figure*}

\vbox{}
\noindent\textbf{Triply resonant cavity magnomechanics}\\
The great flexibility of our system leads to
tremendous advantages. For instance, it allows us
to work in the interesting triple-resonance
condition, where both maximum hybridized modes
simultaneously couple with the phonon mode, as
described by
\begin{eqnarray}
\mathcal{H}_{\mathrm{mb}} & = & \frac{1}{2}\hbar
g_{\mathrm{mb}}(\hat{b}+\hat{b}^{\dagger})(\hat{A}_{+}^{\dagger}\hat{A}_{-}+\hat{A}_{-}^{\dagger}\hat{A}_{+}).
\end{eqnarray}
By adjusting the direction of bias field or the
position of the YIG sphere inside the cavity, we
can tune the hybrid mode splitting to match the
phonon frequency $\omega_{\mathrm{b}}$. In this
case, both the drive and probe photons can be
applied on-resonance with the hybrid modes (top
inset of Figs.\,\ref{fig:fig4}A and B), resulting
in a drastically enhanced magnomechanical
coupling. For the red-detuned drive, the
transparency windows at various driving powers
are plotted in Fig.\,\ref{fig:fig4}A. In addition
to the red shift of the center frequency, the
linewidth of the transparency windows exhibits a
clear linear dependence on the driving power
(Fig.\,\ref{fig:fig4}C, red squares). With a
driving power of only 8.0 dBm, the linewidth
increases from its intrinsic value
$0.62\,\mathrm{kHz}$ to $2.12\,\mathrm{kHz}$,
corresponding to a cooperativity $C=2.4$. As a
comparison, a driving power of $34\,\mathrm{dBm}$
is used to achieve the same cooperativity when
the drive is applied off-resonance, indicating
the drastic enhancement of the magnomechanical
interaction induced by the triple-resonance
condition. The reflection signal for the blue
detuning situation is plotted in
Fig.\,\ref{fig:fig4}B at various driving powers.
As the driving power increases, the center
frequency of the small phonon-induced resonance
inside the hybrid mode is blue shifted, and its
linewidth linearly decreases
(Fig.\,\ref{fig:fig4}C, blue circles).

A direct comparison of Figs.\,\ref{fig:fig4}A and
B reveals distinctly different spectral
lineshapes of the phonon-induced resonances. The
same as in the case of off-resonance drive, we
observed MMIT for the red-detuned drive in the
triply resonant system, with the peak height and
linewidth of the transparency window increasing
with the driving power. While for the
blue-detuned drive, we observed the transition
from MMIA to MMIT, and then to MMPA and
eventually self-sustained oscillation as we
increase the driving power. These observations
lead to a unified explanation about the modified
spectral lineshape (which is not limited to the
triple resonance situation): the coupling with
phonon introduces additional dissipation and
phase shift to the hybridized modes and therefore
changes their lineshapes. With the presence of a
parametric drive, the effective dissipation rate
of the hybrid mode is modified from $\kappa$ to
$\kappa(1\pm C)$, which increases for the
red-detuned drive while decreases for
blue-detuned drive. Given a fixed external
coupling rate $\kappa_{\mathrm{e}}$, the
on-resonance reflectivity of the cavity is
\begin{equation}
r=\frac{1\pm
C-2\frac{\kappa_\mathrm{e}}{\kappa}}{1\pm C}.
\end{equation}
Therefore, depending on the external coupling
condition and the driving power, the reflection
spectra lineshape varies among MMIT, MMIA or
MMPA.

The measured on-resonance reflectivity for an
under-coupled hybrid mode agrees well with our
theoretical model (Fig.\,\ref{fig:fig4}D). For
the red-detuned drive, the increasing linewidth
with elevated driving power causes the mode
further under-coupled and therefore the
on-resonance reflectivity increases. On the
contrary, for the blue-detuned situation, the
decreasing linewidth first leads to critical
coupling and then over coupling condition,
yielding a dip in the reflectivity followed by a
rapid increase which diverges as $C\approx1$ at a
driving power of 6.2 dBm. The deviation of the
measured reflectivity from the theoretical
prediction can be attributed to thermal
instability or gain-bandwidth-product limitation,
which also limit the highest measurable
parametric gain to 3 dB. When the hybrid mode is
tuned to over coupled, the increase of the
parametric gain with the driving power is more
gradual, and therefore a much higher parametric
gain up to $25$ dB is achieved before reaching
instability (Fig.\,\ref{fig:fig4}E). The observed
MMPA is similar to the electromechanical
parametric amplifiers \cite{Massel2011_Nature}
but with unprecedented tunability. Further
increasing the driving power leads the system
into the instable regime where the phonon mode
experiences self-sustained oscillation. The
threshold behavior of the measured emission power
from the Stokes sideband, as shown by the inset
of Fig.\,\ref{fig:fig4}D, indicates the onset of
the phonon lasing \cite{Spencer1958}.

\vbox{}
\noindent\textbf{Conclusion}\\
The demonstration of the coherent magnon-phonon
interaction, including the MMIT (MMIA) and MMPA,
provides a versatile platform for the coherent
information processing. Besides, as YIG also
possesses great optical properties such as low
optical loss and optomagnetic nonreciprocity, our
study shows great potential for integrating
different systems, including microwave, optical,
mechanical and magnonic systems, in a single
device and realizing information inter-conversion
among these different information carriers.
Distinguished from opto- or electro-mechanical
systems, our cavity magnomechanical system shows
high level of tunability which allows the
resonance be externally controlled in a wide
frequency range. Moreover, such a complex system
is compatible with superconducting quantum
circuits \cite{Tabuchi2015}. All of these are not
only crucial for realizing long desired functions
such as microwave-to-optical conversion
\cite{Andrews2014_NPhys, Bagci2014, Bochmann2013,
Vitali2012, Clerk2012}, but also provide a
flexible platform that intrigues the fundamental
study of exotic magnetic excitations.

\vbox{} \vbox{}\vbox{}
\noindent \textbf{ACKNOWLEDGEMENTS}\\
\vbox{}\\ \noindent We thank N. Zhu for gluing
the YIG sphere to the silica fiber. This work was
supported by DARPA MTO/MESO program
(N66001-11-1-4114). C.Z., L.J. and H.X.T.
acknowledge support from LPS through
an ARO grant (W911NF-14-1-0563) and Packard Foundation.
L.J. also acknowledges support from the Alfred P. Sloan Foundation.\\

%%%%%%%%%%%%%%%%%%%%%%%%%%%%%%%%%%%%%%%%%%%%%      FIGURES%%%%%%%%%%%%%%%%%%%%%%%%%%%%%%%%%%%%%%%%%%

\global\long\def\figurename{Figure}

%For contributions with methods sections, legends%should not contain any details of methods, or%exceed 100 words (fewer than 500 words in total%for the whole paper). In contributions without%methods sections, legends should be fewer than%300 words (800 words or fewer in total for the%whole paper).

\newpage{}

\newpage{}

\newpage{}

\end{document}